\input epsf.tex
\documentclass[12pt]{article}
\usepackage{psfig,picinpar,floatflt,amssymb,epsf}
\newcommand{\resection}[1]{\setcounter{equation}{0}\section{#1}}
\newcommand{\appsection}{\addtocounter{section}{1} \setcounter{equation}{0}
             \section*{Appendix \Alph{section}}}

\def\ch{\cosh}

\def\vep{\epsilon}

\def\bd{\begin{displaystyle}}
\def\ed{\end{displaystyle}}
\def\ba{\begin{array}}

\def\ea{\end{array}}
\def\EQ{\begin{equation}}
\def\EN{\end{equation}}
\def\bea{\begin{eqnarray}}
\def\eea{\end{eqnarray}}
\def\beano{\begin{eqnarray*}}
\def\eeano{\end{eqnarray*}}

\def\hs{\hspace{0.1in}}

\def\be{\beta}

\begin{document}
\oddsidemargin 5mm
\setcounter{page}{0}
\setcounter{page}{0}
\begin{titlepage}
\begin{flushright}
ISAS/EP/99/23\\
\end{flushright}
\vspace{0.5cm}
\begin{center}
{\large {\bf Bosonic-type $S$-Matrix, Vacuum Instability\\
and CDD Ambiguities}}\footnote{Work done under partial support of 
the EC TMR Programme {\em Integrability, non--perturbative effects 
and symmetry in Quantum Field Theories}, grant FMRX-CT96-0012} \\
\vspace{1.5cm}
{\bf G. Mussardo$^{a,b}$ and P. Simon$^{c}$} \\
\vspace{0.8cm}
$^a${\em Dipartimento di Fisica, Universita' dell'Insubria, Como (Italy)}\\
$^b${\em Istituto Nazionale di Fisica Nucleare, Sezione di Trieste}\\
$^c${\em International School for Advanced Studies, Via Beirut 2-4, 
34013 Trieste, Italy} \\ 
\end{center}
\vspace{6mm}
\begin{abstract}
\noindent
We consider the simplest bosonic--type $S$-matrix which is usually 
regarded as unphysical due to the complex values of the finite volume
ground state energy. While a standard quantum field theory interpretation 
of such a scattering theory is precluded, we argue that the physical 
situation described by this $S$--matrix is of a massive Ising model 
perturbed by a particular set of irrelevant operators. The \\    
presence of these operators drastically affects the stability of the original 
vacuum of the massive Ising model and its ultraviolet properties. 
\end{abstract} 
\vspace{5mm}
\end{titlepage}
\newpage

\setcounter{footnote}{0}
\renewcommand{\thefootnote}{\arabic{footnote}}

\resection{Introduction}

\,

In recent years, remarkable progress has been witnessed 
in the study of the scaling region nearby the fixed points 
of the Renormalization Group (RG) of 2-D statistical models. 
Such progress has been possible thanks to new ideas related
to the $S$--matrix theory \cite{ZZ,Zam} and there is now a 
well--defined program to analyse off--critical statistical 
models. In short, this usually consists in implementing 
the following steps: (a) a deformation of a Conformal 
Field Theory (CFT) describing the critical point by 
means of a {\em relevant} integrable operator which 
drives the model away from criticality; 
(b) a consequent determination of the exact spectrum of the 
massive or massless excitations and the elastic $S$--matrix 
scattering amplitudes thereof; (c) and finally, a reconstruction 
of the off--critical correlators by means of the Form Factor 
Approach also accompanied by an analysis of the finite 
size behaviour of the theory by means of the Thermodynamics 
Bethe Ansatz (TBA). The scattering theory of the Ising model 
in a magnetic field \cite{Zam} together with the calculation 
of its correlators \cite{DM} provide an explicit example of 
a statistical model solved away from criticality along the 
line of the above program. Solutions for many other off--critical 
models have also been found (see for instance [3-19]).
Hereinafter the statistical models 
solved according to an $S$--matrix program will be simply 
referred as {\em bootstrap models}. 

All known boostrap models greatly differ from each 
other by the nature of their spectrum, for the total 
number of particles and for the detailed structure of their 
scattering amplitudes. However, looking at them as a whole, 
they share an important feature, which consists in the fermionic 
nature of their scattering amplitudes. Namely, the $S$-matrix involving 
two identical particles takes the value $-1$ at the threshold 
of their $s$-channel, i.e. $ S_{aa}(0) = -1$ (for the notation 
see below). This condition was proved to be quite important 
in the further study of their off--shell behaviour. In the 
Thermodynamics Bethe Ansatz calculations, for instance, 
fermionic--type $S$-matrices lead to well--defined integral 
equations which -- in all known models -- provide real 
solutions for the ground state energy, in perfect 
agreement with the CFT prediction \cite{ZamTBA,KM}. On the 
other hand, in the Form Factor Approach, fermionic 
$S$--matrices give rise to spectral series of the 
correlation functions with fast convergent properties 
(see for instance, \cite{DM,YLZam,ZamIs,CMpol,DM35}).

Among the set of known solved models the peculiar absence of 
examples relative to an $S$--matrix of bosonic--type, i.e. an 
$S$-matrix which satisfies the condition $ S_{aa}(0) = 1$, is 
indeed intriguing. What is the reason? While a widespread 
suspicion is that such $S$-matrices are unphysical\footnote{
It is worth mentioning that in an example based on the algebraic 
Bethe Ansatz \cite{Korepin}, which is a different approach 
to that of this paper, it was shown that the physical vacuum 
of the theory could only be constructed if the particles were 
of fermionic type.}, a full understanding of their features 
and also of their interpretation are still missing. 

Deeply related to the absence of models with a bosonic--type 
$S$--matrix, there is also the question about the role 
played in the scattering theory by the so--called CDD 
factors. It is well known that they give rise to  
ambiguity in the $S$--matrix of a bootstrap system and 
it seems therefore quite striking that all consistent 
boostrap models have managed to successfully resolve  
this ambiguity.  

The aim of this paper is to address some of the above questions. 
In particular, we will draw some conclusions about the interpretation 
of the physical aspects which occur in an integrable theory based 
on a bosonic--type $S$-matrix. We will consider the
simplest  non--trivial scattering theory of this kind, consisting of one 
self--conjugate particle $A$ of mass $m$, with the two--particle 
$S$-matrix given by   
\EQ
S_{AA}(\beta) = - \frac
{\sinh\beta - i \sin\pi a}
{\sinh\beta + i \sin\pi a} \equiv - f_a(\beta) 
\,\,\,.  
\label{mSh}
\EN 
Unless explicitly stated, $a$ is a real 
parameter which, for the invariance of the $S$--matrix under 
$a \rightarrow 1-a$, can be taken in the interval $[0,1/2]$.
In eq.\,(\ref{mSh}) $\beta \equiv \beta_1 -\beta_2$,
where $\beta_i$ is the rapidity variable of each particle entering 
its relativistic dispersion relation $E_i = m \cosh\beta_i$, 
$p_i = m \sinh\beta_i$. Since the Mandelstam variable $s$ is 
given by $ s = (p_1 + p_2)^2 = 2 m^2 (1 + \cosh\beta) $, 
the threshold in this channel is reached when $\beta=0$ and 
for the $S$-matrix (\ref{mSh}) we have 
\EQ
S_{AA}(0) = \left\{
\begin{array}{cl}
1 & \mbox{if $a \neq 0$} \,\,\, ;\\
-1 & \mbox {if $a=0$} \,\,\, ,
\end{array}
\right.
\label{bosonic}
\EN 
i.e. a bosonic--type $S$-matrix except when $a=0$. Notice 
that the asymptotical behaviour of this $S$--matrix is given 
by $\lim_{\beta \rightarrow \infty} S(\beta) = -1$. 
As it will become clear in the next section, the $S$--matrix 
(\ref{mSh}) can also be equivalently regarded as a pure CDD 
factor added to an initial fermionic--type $S$--matrix. 

A word of caution. In the following we will often use the 
terminology "QFT" to denote briefly and concisely a hypotetical 
theory underlying the scattering processes. The terminology 
does not automatically mean that this underlying theory is 
a {\em consistent} Quantum Field Theory, the peculiar features 
of which are precisely the object of analysis. 

The paper is organised as follows. In Section 2 we briefly 
discuss the CDD factors and their relation with a perturbed CFT. 
In Section 3 we discuss the difficulties which the bosonic--type 
$S$--matrix (\ref{mSh}) pose both in the TBA analysis and in the 
Form Factor Approach -- difficulties which prevents its 
interpretation in terms of a standard quantum field theory. 
Section 4 is devoted to an interpretation of the bosonic--type 
$S$--matrix (\ref{mSh}) as the one coming from the critical 
Ising model perturbed by a combination of relevant and 
irrelevant operators. We will argue that this combination induces an
instability  in the ground state of the theory, in agreement with the 
results of the TBA analysis for the bosonic--type 
$S$--matrix (\ref{mSh}). Finally our conclusions are 
summarised in Section 5.

\resection{CDD Factors and Deformations of CFT}

To analyse the scaling region around a critical point described 
by a CFT, one usually considers the deformation of the conformal 
action ${\cal A}_{CFT}$ by means of one of its {\em relevant} 
operators $\Phi(x)$ 
\EQ
{\cal A} = {\cal A}_{CFT} + \lambda \int \Phi(x) d^2x \,\,\,.
\label{action}
\EN 
The resulting action -- which describes a RG flow from the 
original CFT to another fixed point -- gives rise in the Minkowski 
space to the scattering processes among the massless or 
massive excitations present away from the critical point 
\cite{Zam}. Important simplifications occur when the action 
(\ref{action}) defines an integrable quantum field theory: in this 
case, in fact, all the scattering processes are purely 
elastic and can be expressed in terms of the two-particle 
scattering amplitudes $S_{ab}(\beta)$. On a general basis, 
these amplitudes satisfy the unitarity and crossing symmetry 
conditions\footnote{For simplicity we consider here theories 
with a massive non-degenerate spectrum of self-conjugate 
particles.} 
\EQ
\begin{array}{l}
S_{ab}(\beta) S_{ab}(-\beta) = 1 \,\,\, ;\\
S_{ab}(\beta) = S_{ab}(i \pi -\beta) \,\,\,.
\end{array}
\label{unitcros}
\EN 
Possible bound states which occur in the scattering processes 
are promoted to asymptotic states so that the amplitudes satisfy 
the additional bootstrap equations 
\EQ
S_{cd}(\beta) = S_{bd}(\beta -i \bar{u}_{bc}^a) 
S_{ad}(\beta + i \bar{u}_{ac}^b) \,\,\, ,
\label{bootstrap}
\EN      
where $\bar{u} \equiv \pi - u$ and $i u_{ab}^c$ is the location 
of the simple pole in $S_{ab}(\beta)$ which identifies the bound 
state $A_c$ in the scattering channel $A_a A_b \rightarrow A_a A_b$. 

The solution of the above equations presents the well--known 
ambiguity relative to the CDD factors. Let $\hat S_{ab}(\beta)$ be 
a set of functions with a minimum number of poles which solve both 
eqs.\,(\ref{unitcros}) and (\ref{bootstrap}). Another solution of 
the first equations (\ref{unitcros}) can easily be obtained by 
multiplying each amplitude $\hat S_{ab}(\beta)$ by the so--called 
CDD factors, i.e. an arbitrary product of functions $f_a(\beta)$ 
defined in eq.\,(\ref{mSh}),  
$ 
\hat S_{ab}(\beta) \longrightarrow {\cal C}_{ab}(\beta) 
\hat S_{ab}(\beta)
$, 
where 
$ 
{\cal C}_{ab}(\beta) = \prod_{a_i}^{N_{ab}} f_{a_i}(\beta)$. 
In fact, each $f_a(\beta)$ automatically satisfies eqs.\,(\ref{unitcros}) 
and hence their product. Additional constraints on such CDD factors 
may come from the bootstrap equations (\ref{bootstrap}), which they 
also have to satisfy. Since the CDD factors do not introduce 
extra poles in the physical sheet $\makebox{$ 0 \leq$ Im $\beta 
\leq \pi$}$, the conclusion is that the knowledge of the
structure of the bound states alone cannot uniquely fix the 
scattering theory. This ambiguity is however not harmless since 
scattering theories which differ by CDD factors have a distinct 
physical content, in particular a different ultraviolet 
behaviour\footnote{A well known example is provided by the 
Affine Toda Field Theories \cite{Dorey,Toda} which share the same 
structure of bound states of some deformed CFT but have different 
central charges \cite{KM}.}. 

How can we interpret $S$--matrices which differ by CDD factors in 
terms of an underlying action? First of all, the CDD factors do not 
spoil -- by construction -- the integrable nature of the original 
$S$--matrix, simply because they only add extra phase--shifts to the 
original elastic scattering amplitudes. Moreover, they also do not alter 
the structure of the bound states -- an infrared property of theory --, 
because they do not introduce extra poles in the $S$--matrix. Since they
mainly influence the ultraviolet behaviour, it seems natural to assume that 
$S$--matrices related to one another by CDD factors are associated 
to integrable actions which, in the vicinity of a fixed point, 
may differ for insertion of irrelevant scalar operators\footnote{
It is worth mentioning that the same line of reasoning has been 
successfully applied in the context of Boundary Quantum Field Theories, 
with the CDD factors entering the reflection scattering matrix put in 
correspondance with the irrelevant boundary deformations of the 
theory, see \cite{Saleur}.} $\eta_i(x)$ 
\EQ
{\cal A} = {\cal A}_{CFT} + \lambda \int \Phi(x)\, d^2x + 
\mu_i \int \eta_i(x) \, d^2x \,\,\,.
\label{actionirr}
\EN 
These irrelevant operators should preserve the set of 
conserved currents (or part of it) of the original action 
(\ref{action}) and they should also share the same symmetry 
properties of the relevant operator $\Phi(x)$. Their presence 
has however several consequences. In fact,from a RG geometric point 
of view the irrelevant operators $\eta_i(x)$ are associated 
to RG trajectories which flow into the fixed point, whereas the 
relevant operators are associated to those RG trajectories 
which depart from it. Hence, the simultaneous presence of 
relevant and irrelevant operators gives rise to a RG trajectory 
which will pass very closely to the fixed point in question but 
without stopping at it (Figure 1). In its vicinity, its action 
can be parameterised as in (\ref{actionirr}). However, in contrast 
with the QFT defined by the action (\ref{action}), the 
action given by eq.\,(\ref{actionirr}) does not automatically 
ensure the consistency of the relative quantum field theory in 
the entire RG space of the couplings, i.e. on all possible distance 
scales. Obviously, there are some preliminary steps in order to 
give a meaning to the expression (\ref{actionirr}). 

First of all, the irrelevant operators initially present in 
(\ref{actionirr}) will be accompanied by an infinite number 
of counterterms, so that the final form of the action of the theory 
will be actually given by the one of (\ref{actionirr}) together 
with this infinite tower of higher irrelevant operators. This 
status of art however is not catastrophic as it may appear since the theory 
can still remain predictive in view of the fact that all counterterms 
can be constrained by the request of integrability \cite{ZamTIM,Lesage}.
A real source of problem is another one. In fact, in contrast to 
those theories which are obtained by perturbing a CFT by means of 
relevant operators, where their ultraviolet behaviour is completly 
under control and ruled in fact by the initially CFT, the ultraviolet 
behaviour of those quantum field theories defined as in (\ref{actionirr}) 
will be, instead, a--priori unknown. Since at the short distances the 
effective coupling constants of the irrelevant operators become 
extremely large, it should not be a surprise to find out that
one may define a consistent quantum field theory in this regime only 
for special sign of the leading irrelevant coupling\footnote{There 
is a simple and even trivial analogy of this situation with 
other quantum field theory problems: let us consider for instance 
the hamiltonian of massive boson $H= \frac{1}{2} \left((\nabla \Phi)^2 + 
m^2 \Phi^2\right) + \lambda \Phi^4$ to which we add the interaction 
$ \mu \Phi^6$. Assume that the solution of the theory at $\mu =0$ 
is exactly known. Although a perturbation theory can be formally applied 
for any sign of the new coupling $\mu$, it is obvious that the new 
hamiltonian will define a consistent theory only for positive 
values of this coupling since for $\mu$ negative, no matter how small it 
may be, there is no longer a stable vacuum.}. This condition 
is realised, for instance, for the roaming trajectory of the Toda 
Field Theories \cite{Roaming} which in each neighborhood of the 
fixed points it passes, may be regarded as a RG flow induced by a 
combination of the relevant operator $\Phi = \varphi_{1,3}$ and the 
irrelevant operator $\eta = \varphi_{3,1}$, for an appropriate set 
of the couplings \cite{Lassig}. On the contrary, another choice of 
the sign of the couplings nearby the critical points may induce an 
ultraviolet instability of the ground state of the theory 
at certain scale $\Lambda$ which will spoil a quantum field theory 
interpretation for distance scales $R \leq \Lambda$. As we will see, 
this seems to be the scenario associated to the $S$--matrix (\ref{mSh}). 

\resection{Difficulties with the bosonic--type $S$--matrix}

A deceptively simple $S$--matrix affected by a CDD ambiguity 
is obtained by multiplying a scattering amplitude $S_{aa}(\beta)$ 
by $-1$ (which may be considered as a particular $f_a$ term, i.e. 
$\lim_{\Delta\rightarrow \infty} f_{\frac{1}{2} + i \Delta}(\beta)$). 
If the original scattering theory does not have any bound states, 
i.e. if there are no constraints coming from the non--linear 
bootstrap equations (\ref{bootstrap}), there is evidently no 
obstacle to this multiplication. Consequently, by means of this 
CDD factor we can swap from a fermionic--type $S$--matrix to a 
bosonic one and viceversa. This is the case, for instance of 
the $S$--matrix of the Sinh-Gordon model, which is given by 
$S^{Sh} = f_B(\beta)$, where $B$ is a function of the coupling 
constant of the model \cite{AFZ}. The result of this multiplication 
is in fact the bosonic--type $S$--matrix (\ref{mSh}). For this 
reason, the model associated to the $S$--matrix (\ref{mSh}) may 
be referred to as ``minus Sinh-Gordon model'' (mShG). 

There is, however, another way of looking at the $S$--matrix 
(\ref{mSh}). It may be seen, in fact, as the one obtained 
by multiplying $S = -1$ by a CDD factor $f_a(\beta)$. 
$S = -1$ is the $S$--matrix of the thermal Ising model 
\cite{Swieca} and therefore the analysis of this  
bosonic--type $S$--matrix may be pursued along the 
considerations presented in the previous section, as will 
be done in Section 4. In the meantime, let us investigate what is 
the nature of the problems which are posed by the bosonic--type 
$S$--matrix (\ref{mSh}), regardless of its interpretation. 
We will initially study its finite size and ultraviolet behaviour by
means of the TBA approach and we will then proceed to the 
investigation of its Form Factors. 
 
\subsection{Thermodynamical Properties} 

There is a simple way to see that the theory defined by the 
$S$-matrix (\ref{bosonic}) presents some problematic aspects 
from the point of view of a quantum field theory. Let us compute 
in fact the free energy $F(mR)$ of such a theory on a cylinder of 
length $L$ and radius $R$ (with $L \gg R$) by means of the TBA 
approach \cite{YY,ZamTBA}. As usual, $R^{-1}$ can be interpreted as 
the temperature of the one--dimensional system living on an 
infinite line, $R^{-1} = T$. As it is well--known, once the free energy
$F(m R)$ has been computed, the ground state Casimir energy $E_0(m R)$  
-- parameterised in terms of the effective central charge 
$C(m R)$ as 
\EQ
E_0(mR) \equiv -\frac{\pi}{6 R} C(m R) \,\,\,,
\label{Casimir}
\EN 
will be given by  
\EQ
E_0(m R) = F(m R) \,\,\,.
\label{association}
\EN

In the TBA approach, the free energy $F(m R)$ is
initially expressed in terms of the functional  
\EQ
f(\rho,\rho^{(r)}) = {\cal E}(\rho^{(r)}) - \frac{1}{R} 
{\cal S}(\rho,\rho^{(r)})  - \mu {\cal N}(\rho^{(r)}) \,\,\,. 
\label{functional}
\EN 
The energy term is given by 
\EQ
{\cal E}(\rho^{(r)}) = \int m \cosh\beta \,\rho^{(r)}(\beta) \,
d\beta \,\,\,,
\label{hamiltonian}
\EN 
the entropy term, in the bosonic case, is given by 
\EQ
{\cal S}(\rho,\rho^{(r)}) = \int d\beta 
\left[(\rho + \rho^{(r)}) \log(\rho + \rho^{(r)}) - 
\rho \log \rho - \rho^{(r)} \log \rho^{(r)} \right] \,\,\,, 
\label{entropy}
\EN 
and the particle number is expressed as 
\EQ
{\cal N}(\rho^{(r)}) = \int \rho^{(r)}(\beta)\,d\beta \,\,\,.
\label{number}
\EN 
In the above formulas, $\rho^{(r)}(\beta)$ is the density of 
occupied states (roots) of rapidity $\beta$ whereas $\rho(\beta)$ 
is the total density of states (roots and holes), $\rho = \rho^{(r)} + 
\rho^{(h)}$. They are related each other by the 
integral equation 
\EQ
\rho(\beta) = \frac{m}{2\pi} \cosh\beta + 
\int \varphi(\beta - \beta') \rho^{(r)}(\beta') \frac{d\beta'}{2\pi} 
\,\,\, , 
\label{integralrelation}
\EN
with the kernel $\varphi(\beta)$ given in our case by 
$$ 
\varphi(\beta) = -i \frac{d}{d\beta} \log S = 
\frac{2 \sin\pi a \cosh\beta}{\sinh^2\beta + \sin^2\pi a} \,\,\,.
$$
Let us assume that the functional (\ref{functional}) admits a minimum 
with respect to the distributions $\rho(\beta)$ and $\rho^{(r)}(\beta)$, 
with eq.\,(\ref{integralrelation}) acting as a constraint. If this would 
be the case, the partition function $Z(L,R)$ in the presence of a chemical 
potential $\mu$ ($z \equiv e^{\mu R}$) and in the limit $L \rightarrow 
\infty$ can be computed in the saddle point approximation and expressed 
as   
\EQ
Z(L,R) \sim \exp\left[- m L \hat F(R)\right] 
\,\,\,\,\, , \,\,\,\,\,  
\hat F(R) = \int_{-\infty}^{+\infty} \cosh\beta \,
\log(1 - z e^{-\epsilon(\beta)}) \,\frac{d\beta}{2\pi} \,\,\, ,
\label{freeenergy}
\EN 
where $\hat F(R)$ is the value of the functional $R f(\rho,\rho^{(r)})$ 
computed at its minimum and the minimum condition is provided by 
the integral equation satisfied by the pseudo-energy $\epsilon(\beta)$ 
\EQ
\epsilon(\beta) = m R \cosh\beta + \int_{-\infty}^{+\infty} 
\varphi(\beta-\beta') \log(1 - z \,e^{-\epsilon(\beta')}) 
\frac{d\beta'}{2\pi} \,\,\,.
\label{pseudoenergy} 
\EN 
In terms of $\epsilon(\beta)$, the distributions $\rho(\beta)$ and 
$\rho^{(r)}(\beta)$ which provide the minimum for the functional 
(\ref{functional}) are related each other by 
\EQ
\rho^{r} = \rho \,\frac{z \,e^{-\epsilon}}{1 - z\, e^{-\epsilon}}\,\,\,.
\label{equilibrium}
\EN 

\vspace{3mm}

The above discussion briefly summarises the basic ideas of the
Thermodynamical Bethe Ansatz approach. Let us now analyse how they 
are actually implemented in the case of our bosonic model, starting our
analysis from the expression of the free--energy obtained in the saddle 
point approximation of the TBA equations, i.e. from eqs. (\ref{freeenergy}) 
and (\ref{pseudoenergy}). At zero chemical potential $z =1$, no matter 
how small $a$ may be (but never zero), the numerical solution of the 
above equations shows the existence of a critical value of the variable 
$R$, such that for $R < R_c(a)$, the free energy assumes complex values 
(Figure 2). It is worth noting the novelty of this situation, which in 
fact does not occur in all other known (fermionic) models, analysed in 
the past in terms of the TBA equations (see, for instance, \cite{KM}). 
How can we interpret the occurrence of this critical point in $R$ and the 
complex values assumed by the free energy for $R < R_c$? What are 
the consequences of this result? The answer to these questions 
may be expressed both in physical or in purely mathematical terms. 

The simplest physical answer is the following. Since the TBA equations 
provides the full resummation  of the virial expansion of $Z(L,R)$,
$R^{-1}_c(a)$ is then identified with the radius of convergence of the 
corresponding series. The existence of this singularity indicates 
a non-trivial ultraviolet behaviour of the theory, in particular, it 
definetly precludes its standard interpretation in terms of a Conformal 
Field Theory deformed by a relevant operator\footnote{
The problem with the ultraviolet behaviour of such theory can 
also be seen without having recourse to numerical integration 
of the above equations. In fact, in consistent theories, the 
ultraviolet behaviour is ruled by the so--called kink configuration 
of eq.\,(\ref{pseudoenergy}) and the corresponding dilogarithmic 
functions. In our case the plateau of the kink configuration, 
solution of algebraic equation $\epsilon_{pl} = \log (1 - 
e^{-\epsilon_{pl}})$ is in fact complex, $\epsilon_{pl} = 
\pm i \frac{\pi}{3}$.}. 

The singularity at $R = R_c(a)$ may be effectively regarded 
as a sort of Bose--Einstein condensation phenomenon. To show 
this, firstly notice that the partition function (\ref{freeenergy}) 
admits an interpretation as that of a {\em free} bosonic gas 
of exitations $|\chi(\beta)\rangle$ but with dispersion 
relations which depend on the temperature \cite{YY}. In 
particular, for the energy ${\cal E}(\beta)$ of one of these 
excitations $|\chi(\beta)\rangle$ we have ${\cal E}(\beta) \equiv 
\epsilon(\beta)/R$, where $\epsilon(\beta)$ is solution 
of the integral equation (\ref{pseudoenergy}). Hence, in a 
way completely similar to the fermionic case analysed in 
\cite{LCM}, one can establish the equivalence between the 
partition function obtained in eq.\,(\ref{freeenergy}) and the 
one computed in terms of the following series 
\EQ
Z(L,R) = \sum_{n=0}^\infty 
z^n \, \langle \chi(\beta_n) \ldots \chi(\beta_1) |
\chi(\beta_1) \ldots \chi(\beta_n) \rangle e^{-(\epsilon(\beta_1) + 
\cdots + \epsilon(\beta_n))} \,\,\, ,
\label{bosonicsum}
\EN 
where the scalar product of the states is computed according to 
the commutation rules of {\em free bosonic particles}, with a 
regularization given by 
\EQ
\left[ \delta (\beta - \beta' ) \right]^2 \equiv 
\frac{mL}{2\pi} \,\cosh (\beta ) \, \delta (\beta - \beta' ) \,\,\,.
\EN 
In fact, for the partition function in (\ref{freeenergy}) we have 
the following expansion in powers of $(m L)$  
\begin{equation}
Z(L,R) = 1 - (m L) F(R) + \frac{(m L)^2}{2!} (F(R))^2 + \cdots   
(-1)^n \frac{(m L)^n}{n!} (F(R))^n + \cdots
\label{series1}
\end{equation}
where $F(R)$ admits the representation 
\begin{equation}
F(R) = -\sum_{n=0}^{\infty} \frac{z^n}{n} \,{\cal I}_n(R) \,\,\, ,
\label{expansionfree}
\end{equation}
with 
\begin{equation}
{\cal I}_n(R) \equiv  \int \frac{d\beta}{2\pi} \ch \beta 
\,e^{-n \vep (\beta)}
\,\,\,. 
\end{equation}
On the other hand, computing the partition function by using the 
expression (\ref{bosonicsum}), 
\begin{equation}
Z(L,R) = 1 + z {\cal Z}_1 + z^2 {\cal Z}_2 + \cdots z^n {\cal Z}_n + \cdots
\label{series22}
\end{equation}
for the first term we have 
\begin{eqnarray}
{\cal Z}_1 &=& \int \frac{d\beta}{2\pi} \langle 
\beta | \beta\rangle e^{-\vep(\beta)} = \int \frac{d\beta}{2\pi} d\beta'
\delta(\beta - \beta' ) \langle \beta' |  \beta \rangle
e^{-\vep (\beta) } =
\label{2.16}
\\
\,\,\,&\,\,\, = & mL \int \frac{d\beta}{2\pi} \ch (\beta) 
\,e^{-\vep (\beta )} = (m L) \, I_1  \,\,\, ;
\nonumber
\end{eqnarray}
and similarly 
\begin{eqnarray}
{\cal Z}_2 & = &
\frac{1}{2} (m L)^2 I_1^2 + \frac{1}{2} (m L) I_2 \,\,\, ; \nonumber \\
{\cal Z}_3 & = & \frac{(m L)^3}{3!} I_1^4 + \frac{(m L)^2}{2} I_1 I_2 +
\frac{(m L)}{3} I_3 \,\,\, , \\
{\cal Z}_4 & =& \frac{(m L)^4}{4!} I_1^4 + (m L)^3 I_1^2 I_2 +
\frac{(M L)^2}{2} \left[\frac{2}{3} I_1 I_3 + \left(\frac{I_2}{2}\right)^2
\right] +\frac{(m L)}{4} I_4 \,\,\, ,\nonumber 
\end{eqnarray}
etc. Hence, it is easy to see that the series (\ref{series1}) precisely
coincides with that of eq.\,(\ref{series22}), the only difference being 
in the arrangement of the single terms: the terms proportional 
to $(m L)$ in all $Z_n$ gives $F(R)$ as their sum, whereas the sum 
of the terms proportional to $(m L)^n$ appearing in all $Z_n$ gives 
rise to the higher power $(F(R))^n$.

In the light of these considerations, it is then natural to expect
that the partition function (\ref{freeenergy}) will have a singularity 
at that value of the temperature $R^{-1}_c(a)$ where the energy 
of the excitation (at zero rapidity) vanishes (Figure 3). When this 
happens, the contribution coming from the average occupation number 
with ${\cal E}=0$ becomes as important as the entire series, producing 
a singularity in the free energy which resembles that of the usual 
Bose--Einstein condensation. One should not be surprised that such 
effect occurs in this low dimensional system due to the entanglement of 
the energy ${\cal E}(\beta)$ of its excitation with the temperature 
itself of the system. In conclusion, from the TBA analysis one infers 
that in our bosonic--type model there is a scale of distance 
$\Lambda = R_c$ below which there is a change in the ground 
state of the theory. Obviously it is not surprising to find out  
that the evaluation of the series beyond its radius of convergence  
provides complex values, although no direct physical meaning can be 
assigned to them. 
   
From a purely mathematical point of view, the occurrence of complex 
values in $\hat F(R)$ simply indicates the {\em failure} of the 
minimization procedure applied to evaluate the free--energy
(\ref{functional}). Namely, in our case, it is no longer true 
that the functional (\ref{functional}) admits a minimum with 
respect to the (positive) real distributions $\rho(\beta)$ 
and $\rho^{(r)}(\beta)$ for {\em any} value of $R$. As a matter 
of fact, the real minima of the functional (\ref{functional}) 
disappear for $R \leq R_c$, becoming complex. Hence the value 
of the functional (\ref{functional}) {\em computed at its 
minima}, i.e. $R \hat F(R)$, takes for $R < R_c$ complex values  
whereas the functional $f(\rho,\rho^{(r)})$ itself becomes 
instead unbounded from below for this range of $R$. This 
implies that the saddle point procedure employed by the TBA 
is no longer justified\footnote{This circumstance is analysed 
in full detail in Appendix A for simplified version of a TBA 
system based on a particular bosonic $S$--matrix.}. However, if 
we still insist on enforcing the validity of the relationship 
(\ref{association}) between the free--energy and the Casimir 
energy of the theory, we should conclude that the divergence 
of the free--energy for $R \leq R_c$ implies the corresponding 
divergence of the central charge of the theory. As we will show 
in the next section, the same conclusion is also reached by 
using the Form Factor approach.

\subsection{Form Factors}

Let us now analyse the computation of the Form Factors associated 
to the $S$--matrix (\ref{mSh}). As we will see, this calculation 
presents some new and distinct features with respects the 
analogous calculation for fermionic--type theories. 

The definition of the Form Factors of a local hermitian scalar 
operator ${\cal O}$ is given as usual by 
\EQ
F^{\cal O}(\beta_1,\ldots,\beta_n) = \langle 0 |
{\cal O} | A(\beta_1), \ldots , A(\beta_n) \rangle \,\,\, .
\label{FF}
\EN 
These quantities satisfy a set of functional equations 
\cite{KW,Smirnov}
\begin{eqnarray}
&& F_n^{\cal O}(\be_1, \dots ,\be_i, \be_{i+1}, 
\dots, \be_n) =
F_n^{\cal O}(\be_1,\dots,\be_{i+1}
,\be_i ,\dots, \be_n) S(\beta_i-\beta_{i+1}) \,\, ,
\nonumber \\
&& F_n(\be_1+2 \pi i, \dots, \be_{n-1}, \be_n ) 
= F_n(\be_2 ,\ldots,\be_n,
\be_1) = \prod_{i=2}^{n} S(\beta_i-\beta_1) 
F_n(\be_1, \dots, \be_n)
\,\,\,\, 
\label{functionaleqs}
\\ 
&&\lim_{\tilde\beta \rightarrow \beta}
(\tilde\beta - \beta)
F_{n+2}^{\cal O}(\tilde\beta+i\pi,\beta,\beta_1,
\beta_2,\ldots,\beta_n) 
 = i\, 
\left(1-\prod_{i=1}^n S(\beta-\beta_i)\right)\,
F_n^{\cal O} (\beta_1,\ldots,\beta_n) \, ,  \nonumber 
\end{eqnarray}
which closely resemble those of the Sinh--Gordon theory \cite{FMS}. 

For scalar operators, the FF depend solely on the 
rapidity differences $\beta_{ij} = \beta_i - \beta_j$. 
As it is evident from the residue equation in 
(\ref{functionaleqs}), the chain of FF with even 
and odd number of particles are decoupled from each other. 
For simplicity, in the sequel we will consider only FF 
of even operators under the $Z_2$ symmetry of 
the model (i.e. those which have only non--vanishing FF on an 
even number of external states). They can be correspondingly 
parameterised as 
\EQ
F_n^{\cal O}(\beta_1,\ldots,\beta_n)\,=\, H_n\, \frac{Q_n(x_1,
\ldots,x_n)}{(\sigma_n)^\frac{n}{2}}\,
\prod_{i<j} \frac{F_{\rm min}(\beta_{ij})}{x_i+x_j}
\,\,\, ,\label{para}
\EN
where $x_i=e^{\beta_i}$, $H_n$ is a normalization 
constant and $F_{\rm min}(\beta)$ is a function 
which satisfies the equations 
\EQ
\begin{array}{ccl}
F_{\rm min}(\beta)&=& S(\beta)\, F_{\rm min}(-\beta)\,\,\, ,\\
F_{\rm min}(i\pi-\beta)&=&F_{\rm min}(i\pi+\beta)\,\,\, . 
\end{array}
\label{Watson2}
\EN
$Q_n(x_1,\ldots,x_n)$ are symmetric polynomials 
in the variables $x_1,\ldots,x_n$ of total degree 
$t_n = \frac{n}{2} (2 n -1)$. They can be expressed 
in terms of the so--called {\em elementary symmetric 
polynomials} $\sigma_j^{(n)}$ given by\footnote{In the 
following we will discard the upper index $n$ 
in the definition of $\sigma_k$, since the total number 
of variables involved will be clear from the context.}
\EQ
\sigma_j^{(n)} = \sum_{i_1 < i_2 < \ldots < i_j}^n 
x_{i_1} x_{i_2} \ldots x_{i_j} \,\,\,.
\label{elementary}
\EN 
Equivalently, they can be obtained in terms of the generating 
function 
\EQ
\prod_{i=1}^n(x+x_i)\,=\,
\sum_{k=0}^n x^{n-j} \,\sigma_j^{(n)}(x_1,x_2,\ldots,x_n)\,\,\,.
\label{generating}
\EN
Although at a formal level, the above formulas of the 
FF appear identical to those of the Sinh--Gordon 
model (except for the extra factor $(\sigma_n)^{\frac{n}{2}}$ 
in the denominator), a substantial difference between
the two models occurs in the properties of $F_{min}(\beta)$. 
As we will see, this difference has far--reaching consequences. 

The two--particle minimal FF of our bosonic--type model 
may be taken as 
\EQ
F_{\rm min}(\beta,a)\,=\,\exp\,\left[-2
\int_0^{\infty} \frac{dx}{x}\frac{ \cosh\left[
\frac{x}{2} (1- 2 a)\right]}{\cosh\frac{x}{2}\sinh x}
\sin^2\left(\frac{x\hat\beta}{2\pi}\right)\right] 
\,\,\, , 
\label{integral}
\EN
which can also be expressed as 
\EQ
F_{\rm min}(\beta,a)\,=\,{\cal N}(a)
\prod_{k=0}^{\infty}
\left|
\frac{\Gamma\left(k+\frac{1}{2}+\frac{a}{2} +
\frac{i\hat\beta}{2\pi}\right)
\Gamma\left(k+1-\frac{a}{2}+\frac{i\hat\beta}
{2\pi}\right)}{\Gamma\left(k+\frac{3}{2}-
\frac{a}{2}+\frac{i\hat\beta}{2\pi}\right)
\Gamma\left(k+1+\frac{a}{2}+\frac{i\hat\beta}
{2\pi}\right)}
\right|^2 \,\,\, ,
\EN
where $\hat\beta = i\pi -\beta$. With $F_{min}(i\pi,a)=1$,
we have 
\EQ
{\cal N}(a) = \prod_{k=0}^{\infty}\left( 
\frac{
\Gamma\left(k+1+\frac{a}{2}\right) 
\Gamma\left(k+ \frac{3}{2} -\frac{a}{2}\right)}
{\Gamma\left(k+1-\frac{a}{2}\right) 
\Gamma\left(k+ \frac{1}{2} + \frac{a}{2}\right)}
\right)^2 \,\,\,.
\EN 
For $a \neq 0$, this function has a {\em finite} value at 
the threshold $\beta=0$ given by 
\EQ
F_{min}(0,a) \equiv {\cal F}(a) = 
\exp\left[\int_0^{\infty} \frac{dx}{x} \frac{\cosh\left[\frac{x}{2} 
(1-2a)\right] \sinh\frac{x}{2}}{\cosh^2\frac{x}{2}} \right] \,\,\,, 
\EN 
which is in stark contrast with the value $F_{min}(0)=0$ of 
fermionic--type theories, which is forced by the condition $S(0)=-1$ 
of their $S$--matrix. The asymptotical behaviour of $F_{min}(\beta,a)$ 
is ruled by 
\EQ
\lim_{\hat\beta \rightarrow \infty} F_{min}(\beta,a) 
\simeq i\, \sqrt{2 \sin\pi a {\cal F}(a)} \,
\exp\left[-\frac{\hat\beta}{2}\right] \,\,\, .
\label{asymptotic}
\EN 
Finally, in the limit $a\rightarrow 0$, $F_{min}(\beta,a)$ 
reduces to 
\EQ
F_{min}(\beta,0) = \frac{i}{\sinh\frac{\beta}{2}} \,\,\, .
\label{limita0}
\EN 
In order to implement the recursive equations associated to the last 
equation in (\ref{functionaleqs}) it is useful to consider the functional 
equation\footnote{Plugging $\beta =0$ in (\ref{shift}), one obtains the 
relation 
$
{\cal F}(a) = \frac{2 \pi^2 {\cal N}^2(a)}{\sin\pi a}
$.}
\EQ
F_{min}(\beta + i \pi,a) F_{min}(\beta,a) = 
\frac{2 i \pi^2 {\cal N}^2(a)}{\sinh\beta + i \sin \pi a} \,\,\, .
\label{shift}
\EN 
The residue equation in (\ref{functionaleqs}) implies that  
the polynomials $Q_n(x_1,\ldots,x_n)$ in eq.\,(\ref{para}), 
satisfy the recursive equations 
\EQ
Q_{n+2}(-x,x,x_1,\ldots,x_n) = x^3 
{\cal D}(x,x_1,\ldots,x_n) \,Q_{n}(x_1,\ldots,x_n) \,\,\, ,
\label{n+2}
\EN 
where 
\EQ
{\cal D}(x,x_1,\ldots,x_n) = 2\,(-1)^{1-\frac{n}{2}} \, 
\sum_{\{l,k,p,q\}=0}^n (-1)^{l+q} x^{4 n -l-k-p-q} 
\frac{\sin[\pi a (p-q)]} {\sin\pi a} \sigma_l^{(n)}
\sigma_k^{(n)}
\sigma_q^{(n)}
\sigma_p^{(n)}
\,\,\,.
\label{functionD}
\EN 
In writing eq.\,(\ref{n+2}) we have posed for convenience 
\EQ
H_{n+2} = H_n \frac{\sin\pi a}
{(2 i \sin\pi a \,{\cal F}(a))^n} \,\,\, .
\label{recnormalization}
\EN
Since ${\cal D}(x,x_1,\ldots,x_n)$ contains four elementary 
symmetric polynomials and each of them is linear in the 
individual variables $x_i$, the partial degrees $p_n$ 
of the polynomials $Q_n$ satisfies the condition $p_{n+2} 
\leq 4 + p_n$. 

After these general considerations, let us now attempt to 
solve the above recursive equations for a particular but 
significant field. Namely, let us assume that there is 
an operator in the theory which plays the role of the 
trace $\Theta(x)$ of the stress--energy tensor $T_{\mu\nu}(x)$. 
If such an operator exists, its Form Factors have some 
distinguished properties which may be useful in their 
explicit determination. First of all, its normalization 
will be given by  
\EQ
F^{\Theta}(i \pi) = \langle A(\beta) |\Theta(0)| 
A(\beta) \rangle = 2 \pi m^2 \,\,\,. 
\label{normalization}
\EN 
Secondly, from the conservation law $\partial^{\mu} 
T_{\mu\nu}(x) =0$, the polynomials $Q_n(x_1,\ldots,x_n)$ 
can be factorised as $Q_n(x_1,\ldots,x_n) = \sigma_1 
\sigma_{n-1} P_{n}(x_1,\ldots,x_n)$. Moreover, assuming 
that this operator is even under the $Z_2$ symmetry of 
the model, its only non--vanishing FF will be those with 
an even number of external particles. Finally, 
the two--point correlation function of $\Theta(x)$ enters 
a sum--rule which permits evaluation of the difference of 
the central charges of the theory by going from the large 
to the short distance scales \cite{cth}
\EQ
\Delta C = C_{uv} - C_{ir} = \frac{3}{4\pi} \int d^2x \,|x|^2\,
\langle 0 |\Theta(x) \Theta(0) | 0 \rangle \, = 
\int_{0}^{\infty} d\mu \,c_1(\mu) \,\,\, ,
\label{ctheorem}
\EN 
where $c_1(\mu)$ may be directly expressed in terms of 
the FF of $\Theta(x)$
\begin{eqnarray}
c_1(\mu)& =&\frac{12}{\mu^3} \sum_{k=1}^{\infty} 
\frac{1}{(2n)!} \int\frac{d\beta_1\ldots 
d\beta_{2k}}{(2\pi)^{2n}}\,\mid F_{2k}^{\Theta}
(\beta_1,\ldots, \beta_{2k})\mid^2 
\label{seriescth}\\
& & \,\,\, \times \,
\delta(\sum_i m\sinh\beta_i)\,
\delta(\sum_i m\cosh\beta_i-\mu)\,\,\, .
\nonumber 
\end{eqnarray} 
Since at the large distance scale the theory is massive, we 
will have $C_{ir}=0$, and the above sum--rule should provide 
a determination of the central charge $C_{uv}$ of the theory in 
its ultraviolet regime.  The first approximation of the above 
sum--rule is obtained by the two--particle contribution 
\EQ
\Delta C^{(2)}(a) \,=\,\frac{3}{2}
\,\int_{0}^{\infty} \frac{d\beta}{\cosh^4\beta} \,
|F_{\rm min}(2\beta,a)|^2
\,\,\, .
\label{twoparticlecth}
\EN
With this information on the FF of $\Theta(x)$, let us 
proceed in the computation of its first representatives. 
Since in the limit $a\rightarrow 0$, our bosonic--type 
$S$--matrix reduces to the one of thermal Ising model 
-- for which the only non--vanishing FF of $\Theta(x)$ 
is the one relative to the two--particle state 
\EQ
F_{min}^{\rm Ising}(\beta) = - 
2 \pi m^2 i \sinh\frac{\beta}{2} \,\,\, ,
\label{isingff}
\EN 
we are forced to take as $Q_2(x_1,x_2)$ 
\EQ
Q_{2}(x_1,x_2) = \sigma_1 (\sigma_1^2 -4 \sigma_2) \,\,\, ,
\label{seed}
\EN 
and $H_2 = -\frac{\pi}{2} m^2$. These two quantities play the 
role of initial values for the recursive equations (\ref{n+2}). 

Notice that by inserting the two--particle FF in 
(\ref{twoparticlecth}), the corresponding quantity 
is an {\em increasing} function of the parameter $a$, 
which takes its minimum value $\Delta C^{(2)} =\frac{1}{2}$ at $a=0$ 
(as in the thermal Ising model) and reaches its maximum 
$\Delta C^{(2)} = 0.876...$ at $a= \frac{1}{2}$ (see Figure 4). 
This increasing monotonic behaviour of $\Delta C^{(2)}(a)$ 
is in contrast with the decreasing monotonic  
behaviour presented by the same quantity for fermionic--type 
theories (see for instance the Sinh-Gordon model \cite{FMS}). 

Using now the factorised form $Q_4(x_1,\ldots,x_4) = 
\sigma_1 \sigma_3 P_4(x_1,\ldots,x_4)$ and the above 
expression for $Q_2(x_1,x_2)$, the recursive equation 
satisfied by $P_4(x_1,\ldots,x_4)$ becomes 
\EQ
P_4(-x,x,x_1,x_2) = -4 
\left[x^8 - x^6 \sigma_1^2 + x^6 \sigma_2 + 
x^4 \sigma_1^2 \sigma_2 -x^4 \sigma_2^2 - 
x^2 \sigma_2^3) \right] (\sigma_1^2 - 4\sigma_2) \,\,\,.
\label{4FF}
\EN 
In the linear space of the symmetric polynomials with 
total degree $t =10$ and partial degree $p \leq 5$, 
the above recursive equation {\em does not} uniquely fix 
the polynomial $P_4(x_1,\ldots,x_4)$, which in fact admits 
a three--parameter family of solutions 
\begin{eqnarray}
P_4(x_1,\ldots,x_4) & = & A_1 (\sigma_1^3 \sigma_3 \sigma_4 
+ \sigma_1 \sigma_3^3 -\sigma_1^2 \sigma_2 \sigma_3^2) + 
A_2 (\sigma_1^2 \sigma_2^2 \sigma_4 + \sigma_2^2 
\sigma_3^2 - \sigma_1 \sigma_2^3 \sigma_3) + \nonumber \\
&&  + A_3 (\sigma_1^2 \sigma_4^2 + \sigma_3^2 \sigma_4 
- \sigma_1 \sigma_2 \sigma_3 \sigma_4) + 32 \,
\sigma_1 \sigma_2 \sigma_3 \sigma_4 + 
\label{4FFsemifinal} \\ 
&& + 16 \sigma_2^3 \sigma_4 -4 \sigma_1^2 \sigma_2 \sigma_3^2 
-4 \sigma_1 \sigma_2^3 \sigma_3 -64 \sigma_2 \sigma_4^2 \,\,\, .
\nonumber 
\end{eqnarray}
This arbitrariness in the four--particle FF of $\Theta(x)$ may 
be partially reduced by imposing an additional condition on 
its FF, namely that they fulfill the ``cluster equations" 
\cite{KM,DSC,AMV}  
\EQ
\lim_{\Lambda \rightarrow \infty} 
F_4^{\Theta}(\beta_1 + \Lambda,\beta_2 + \Lambda,\beta_3,\beta_4) 
=\frac{1}{\langle \Theta \rangle} \,
F_{2}^{\Theta}(\beta_1,\beta_2) \, F_{2}^{\Theta}(\beta_3,\beta_4)
\,\,\, . 
\label{cluster}
\EN 
By using the cluster properties of the elementary symmetric 
polynomials $\sigma_k$, as determined in \cite{KM}, together 
with the asymptotical behaviour eq.\,(\ref{asymptotic}) 
of $F_{min}(\beta,a)$ and the vacuum expectation value 
$\langle \Theta \rangle = \frac{\pi m^2}{2 \sin\pi a}$, the 
cluster equation (\ref{cluster}) provides the following 
conditions on the constants $A_1$ and $A_2$ 
\EQ
A_1 = -1 \,\,\,\,\, , \,\,\,\,\, 
A_2 = -4 \,\,\,.
\label{extraconditions}
\EN 
However, the constant $A_3$ cannot be fixed in this way, since all 
terms proportional to it in (\ref{4FFsemifinal}) are sub--leading
in the above limit (\ref{cluster}). This arbitrariness 
of the FF of $\Theta(x)$, which persists for higher FF, 
shows that in the mShG model the $S$--matrix alone 
cannot uniquely fix the matrix elements of one of its significant 
operators. This arbitrariness in the FF of $\Theta(x)$ should 
be contrasted with their {\em unique} determination in all 
known examples of fermionic--type $S$--matrix\footnote{It is  
easy to check that also higher Form Factors of $Z_2$ odd operators (as  
the one which creates the particle $A$ of the mShG model) 
cannot uniquely fixed.} (see, for instance \cite{DM,YLZam,DM35,FMS}). 

Concerning the $c$--theorem sum rule, some conclusions  
can be also drawn despite the arbitrariness present in the 
FF of $\Theta(x)$. Namely, let us assume that the arbitrary 
constants present at each stage of the recursive equations 
could be fixed according to some additional principle. 
What will then be the {\em generic} behaviour of the series
(\ref{seriescth}) entering the $c$--theorem? For the finiteness 
of the value assumed by the FF at all particle thresholds, 
an estimate of the integral (\ref{ctheorem}) can be provided 
by only considering the contributions at all thresholds. 
To do so, let us first observe that the sum (\ref{seriescth}) 
--  apart from the prefactor $12/\mu^3$ --, is an integral of the square 
of each FF integrated on the phase--space $\Omega_{2k}(\mu)$ 
of $2k$ particle, defined as 
\EQ
\Omega_{2k}(\mu) = \int 
\frac{d p_1}{(2 \pi) 2 E_1} 
\cdots 
\frac{d p_{2k}}{(2 \pi) 2 E_{2k}} \delta(\mu - E_1 - \cdots -E_{2k}) \,
\delta(p_1 + \cdots p_{2k}) \,\,\,.
\label{defphasespace}
\EN  
Near the threshold this quantity may be expressed as  
\EQ
\Omega_{2 k}(\mu) \simeq \frac{1}{\sqrt{4 \pi k m}} 
\,\left(\frac{1}{8 \pi m}\right)^k \,
\frac{1}{\Gamma\left(k-\frac{1}{2}\right)} \, 
\left(\mu - 2 k m\right)^{k - \frac{3}{2}} \,\,\, .
\label{phasespace}
\EN 
Concerning the value of the $F_{2k}(\beta_1,\ldots,\beta_{2k})$
at their threshold ($\beta_{ij} =0$, 
for all $i$ and $j$), we can split it into two terms: the first
one is the product of $H_{2k}$ together with all $F_{min}(0)$, 
given by 
\EQ
H_2 \frac{(\sin \pi a)^{k-1}}
{(2 i \sin\pi a \,{\cal F}(a))^{k (k-1)}} \,
\left({\cal F}(a)\right)^{k (2 k-1)}
= -
\frac{\pi m^2}{2\sin \pi a} \left( 2 i \sin^2\pi a\right)^k \,
\left(\frac{{\cal F}(a)}{2 i \sin \pi a}\right)^{k^2} 
\,\,\,.
\label{valueorigin}
\EN
For the remaining term, we can pose 
\EQ
\left|\frac{Q_{2n}(x_1,\ldots,x_n)}{\sigma_{2n}^n 
\prod_{i<j} (x_i+x_j)}\right|_{\beta_{ij}=0} 
\simeq \left(\frac{\xi}{2}\right)^{k (2k - 1)} \,\,\,. 
\label{estimate}
\EN 
An estimate of the ratio $\rho = \frac{\xi}{2}$ may be 
obtained as follows. From eq.\,(\ref{generating}), 
the numerical values assumed by the elementary symmetric 
polynomials $\sigma_j$ at $\beta_i=0$  
(denoted by $\hat \sigma_j$) is given by the binomial 
coefficient $\hat\sigma_j = \left(\begin{array}{c} 2 k 
\\j \end{array} \right)$. The maximum of these values is 
for $\sigma_k =  \left(\begin{array}{c} 2 k 
\\k\end{array} \right) \simeq 4^k$. At the threshold, in 
the space of symmetric polynomials of total degree $k 
(4 k -1)$ and partial degree $2 (2 k -1)$, the term which 
is expected to dominate in the limit $k\rightarrow 
\infty$ is given by $[(\hat \sigma_k)^{4 k -3} 
\hat\sigma_{2 k}] \simeq 16^{2 k^2 - \frac{3}{2} k}$, 
so that $\xi \sim 16$ and the ratio $\rho$ is therefore 
expected to be always larger than 1. 

Let us now apply the above considerations to 
the $c$--theorem sum--rule. We will approximate the integral 
(\ref{ctheorem}) by using the ``mean theorem'', applied in 
each interval between all successive thresholds. Apart from 
some constants, the series entering eq.\,(\ref{ctheorem}) 
has the following behaviour\footnote{Eq.\,(\ref{approx}) holds for 
$a \neq 0$. For $a = 0$, the $c$--theorem gives $\Delta C = \frac{1}{2}$, 
as in the massive Ising model.}   
\EQ
\Delta C \simeq \sum_{k} \frac{X^{2 k^2}(a) 
Z^{2k}(a)}{(2 k)! \Gamma\left(k-\frac{1}{2}
\right) k} \,\,\, ,
\label{approx}
\EN 
where 
\begin{eqnarray}
&& X(a) = \frac{{\cal F}(a)}{2 \sin\pi a} 
\rho^2 
\,\,\, , \\
&& Z(a) = \frac{2 \sin^2\pi a}{\rho \sqrt{\pi}} 
\,\,\, . \nonumber 
\end{eqnarray}
The nature of the series (\ref{approx}) is obviously 
controlled by the parameter $X(a)$: since the ratio 
$\frac{{\cal F}(a)}{2 \sin\pi a}$ entering $X(a)$ is always 
larger than $1.6..$ ( see Figure 5) and $\rho$ is expected to be larger 
than $1$, we conclude that, due to the finite value assumed 
by the FF at the higher thresholds, the series (\ref{approx}) 
is always divergent. This divergence should be regarded as a 
manifestation of the pathological aspect of the 
quantum field theory associated to mShG model in its
ultraviolet region, as we already learnt 
by the TBA approach. 

\resection{Irrelevant Deformations of the Ising Model}  

With the above insights to the nature of the problems 
presented by the bosonic--type $S$--matrix (\ref{mSh}), 
let us come back to the considerations of Section 2 with 
the aim to identify an underlying action for such an 
$S$--matrix. We will take the point of view in which 
$f_a(\beta)$ in eq.\,(\ref{mSh}) is seen as a CDD factor 
for the fermionic $S$--matrix $S=-1$. The latter is 
identified as the one of the thermal Ising model. Hence, 
we have to look for irrelevant operators in this model which 
both preserve its original integrability and its $Z_2$ 
symmetry such that the extra phase--shift which they  
induce matches with the CDD factor $f_a(\beta)$. 

In the conformal Ising model there are few operators we can 
play with: in fact, we have the operators of the conformal 
family of the identity {\bf I} (which includes the 
stress--energy tensor T), those of the conformal family 
of the energy field ${\cal E}$ and finally those of the 
magnetization field $\sigma$. The operators of the even 
sector of this model, i.e. those of identity and energy 
families, can be written in terms of local expressions of 
the chiral $\Psi(z)$ and anti--chiral $\bar\Psi(\bar z)$ components 
of a Majorana fermionic field. In particular we have 
$E(z,\bar z) = i \bar\Psi(\bar z) \Psi(z)$ and, for the 
analytic and anti--analytic part of the stress energy tensor,  
$T(z) = \frac{1}{2} :\Psi \partial \Psi:$ and $\bar T = 
\frac{1}{2} :\bar\Psi \bar\partial \bar\Psi:$ . 

The original fermionic--type $S$--matrix $S=-1$ originates from  
the CFT action perturbed by the relevant operator of 
the energy field  
\EQ
{\cal A} = {\cal A}_{CFT} + m \int {\cal E}(x) d^2x \,\,\,.
\label{action1}
\EN 
In the euclidean space it may be written as  
\EQ
{\cal A} = \frac{1}{2\pi} \int (\Psi \bar\partial \Psi+ 
\bar\Psi \partial \bar\Psi  + 2 i m \bar\Psi 
\Psi)\, d^2x \,\,\, . 
\label{Majorana} 
\EN 
Let us analyse which operators we can add to this action 
in order to have an extra phase--shift in the $S$--matrix. 
None of these operators can be of the magnetic 
field family as they would spoil both its $Z_2$ symmetry 
and its integrability \cite{Musrep,nonint,fonseca}.
They cannot be either descendent fields of the energy field, 
since they are all quadratic in the fermionic field: 
their insertion in the action (\ref{Majorana}) 
will not induce scattering processes among the fermion particles 
but will only change their dispersion relations. 
Hence, the first irrelevant field which can be introduced 
in (\ref{action1}) in order to have extra scattering processes  
is $T\bar T$ . This operator is in fact quartic in the fermion 
fields and the new action becomes 
\EQ
{\cal A} = \frac{1}{2\pi} \int \left(\Psi \bar\partial\Psi + 
\bar\Psi \partial \bar\Psi  + 2 i m  
\bar\Psi \Psi \right)\,d^2x + \frac{g}{\pi^2 m^2} \int 
\Psi 
\partial \Psi \bar\Psi \bar\partial 
\bar\Psi \, d^2x \,+ \cdots \,\,\, \, 
\label{MajoranaTT} 
\EN 
where $g$ is a dimensionless constant. With the insertion 
of this operator, the theory is still integrable at the lowest 
order \cite{ZamRSOS} although ${\cal A}$ in (\ref{MajoranaTT}) 
should be considered at this stage as an effective action. 
The coupling constant $g$ can be related to the parameter $a$ 
entering the $S$--matrix (\ref{mSh}) as follows. Since the operator 
$T \bar T$ is expected to be effective at very high--energy scales, 
this suggests matching the lowest order coming from the perturbation 
theory of (\ref{MajoranaTT}) with the $S$--matrix (\ref{mSh}), 
both computed in their high--energy limit. In this kinematical 
regime the two particles involved in the scattering can be 
effectively seen as left and right movers. Their momenta $p$ and $q$ 
may be parameterised in terms of a new rapidity variable 
$\theta$ as $p = m e^{\theta_1}$, $q=- m e^{-\theta_2}$, 
so that the Mandelstam variable $s$ is expressed in this 
limit as $s = 2 m^2 e^{\theta_{12}}$. The lowest order 
in $g$ of the scattering amplitude in this regime 
may be computed as in \cite{ZamTIM} with 
the result 
\EQ
{\cal S}(s) \simeq -1 + 2 i g \,\frac{s}{m^2} + \cdots   
\label{lowest}
\EN 
Let us now compare this expression with the one from the 
expansion of the $S$--matrix (\ref{mSh}) around the 
point $\beta_{12} = \infty$. By setting 
$1/\sinh\beta_{12} \equiv e^{\theta_{12}}$ as 
a parameterization around $\beta_{12} = \infty$, 
the $S$--matrix (\ref{mSh}) may be written in the vicinity 
of this point as 
\EQ
S(\theta_{12}) = \frac{s + i\frac{m^2}{\sin\pi a}}
{s -i \frac{m^2}{\sin\pi a}} \simeq -1 + 2 i 
\sin\pi a \,\frac{s}{m^2} + \cdots 
\label{perinfinity}
\EN 
so that we have the identification 
\EQ
g \simeq \sin\pi a \simeq a
\label{identification}
\EN 
i.e. $g$ is a {\em positive} quantity. It is important to note 
that the sign of the above coupling constant is {\em opposite} 
of the one relative to the roaming model nearby the Ising fixed 
point \cite{ZamTIM}. This difference in the sign of the coupling 
relative to the first irrelevant operator seems therefore responsible
for the quite distinct ultraviolet behaviour of the two theories: 
the roaming model has in fact a smooth ultraviolet behaviour 
controlled by the nearest tricritical Ising point (with central 
charge $C =\frac{7}{10}$) whereas the QFT associated to our 
bosonic $S$--matrix develops instead an ultraviolet instability. 
This situation evidently resembles the two different behaviours which 
originate from the change of sign in the example of footnote 5.

Let us finally add a few comments on the full hamiltonian 
of the theory. As discussed in Section 2, the action (\ref{MajoranaTT}) 
will also contain higher derivative terms $\Psi \partial^n \Psi 
\bar\Psi \bar \partial^n \bar\Psi$, so that in general we can write  
\begin{eqnarray}
{\cal A} & =& \frac{1}{2\pi} \int \left(\Psi \bar\partial\Psi + 
\bar\Psi \partial \bar\Psi  + 2 i m  
\bar\Psi \Psi \right)\,d^2x + \frac{g}{\pi^2 m^2} \int 
\Psi 
\partial \Psi \bar\Psi \bar\partial 
\bar\Psi \, d^2x \,+  \,\,\, \, \\
& & +\, \frac{g}{\pi^2}\,\,\sum_{n=2}^{\infty} \frac{a_n}{m^{2n}}\,
\int   
\Psi 
\partial^n \Psi \bar\Psi \bar\partial^n \bar\Psi \, d^2x \,\,\,.
\nonumber 
\label{MajoranaTotal} 
\end{eqnarray}
The a--dimensional constants $a_n$ can be in principle determined 
by the integrability condition of the theory, i.e. by matching the 
expansion of the $S$--matrix in power of $s/m^2$ with the Feynman 
graphs which originate from (\ref{MajoranaTotal}). For the presence of 
the infinite higher derivative terms in the action, the corresponding 
hamiltonian (with respect to the Ising fixed point) will be then a 
non--local one. In the Majorana basis\footnote{In this basis
$\gamma^0 = \left(\begin{array}{ll} 0 & i \\
-i & 0 \end{array}\right) $ 
and $\gamma^1 = \left(\begin{array}{ll} 0 & -
i \\ -i & 0 \end{array} \right)$.} 
denoted as before by $\left(\Psi(x,t),\bar \Psi(x,t)\right)$, 
it is convenient to write the Hamiltonian in terms of a  
dimensionless kernel $V(\hat z)$, ($\hat z = m z$), as    
\EQ 
H = \int :\left[-i \Psi \partial_{x} \Psi + i \bar\Psi 
\partial_{x} \bar\Psi + 2 i m \bar \Psi \Psi \right]: dx  
+ g m^2 \int\int :\bar\Psi \Psi :(x)
V(\hat x -\hat y) :\bar \Psi \Psi:(y) 
dx dy
\label{nonlocal} 
\EN 
in such way that the coupling constants of the higher derivative
terms
are nothing but the higher moments of $V(\hat z)$. Therefore, 
once the coefficients $a_n$ were known, the kernel $V(\hat z)$ 
can be recovered in terms of an inverse Mellin transformation. 
The advantage of expressing the hamiltonian in this way is
twofold. First of all, it shows that the hamiltonian is actually 
renormalisable albeit obviously non--local. Secondly, it provides 
an easy argument to show that the non--local term is the one responsable 
for the instability of the theory. This argument is based on the 
simple hypothesis that $\int V(\hat z) d\hat z > 0$. 
Let us consider in fact the Hamiltonian (\ref{nonlocal}) in the 
Hartree--Fock approximation and let us define an order parameter $\Delta =
-i \langle :\bar \Psi \Psi : \rangle$. For field configurations which 
do not vary on $x$, the value of the Hamiltonian (\ref{nonlocal}) for 
unit volume is given by 
\EQ
\frac{H}{L} = m \Delta - g v \Delta^2 \,\,\, , 
\label{HartreeFock}
\EN 
where $v = \int V(\hat z) d\hat z > 0$. When $g=0$, 
we can always subtract from $H/L$ an infinite constant
(due to the Dirac sea of the fermion) such that we can 
restrict to the positive values of $\Delta$ and the 
minimum of $H/L$ is therefore obtained for $\Delta = 0$. 
This subtraction is equivalent to a selection of the vacuum 
state for the free theory. However, when we switch on 
a positive value of $g$, the quantity $H/L$ becomes once 
again unbounded from below, so that there would be an 
instability of the vacuum originally selected by the free 
theory. Hence the above hypothesis permits to understand 
in an easy way the nature of the problem posed by our 
bosonic $S$--matrix, i.e. the instability of the vacuum state of the 
underlying microscopic Hamiltonian. 
 
\resection{Conclusions}

In this paper we have discussed the physical properties 
of the simplest bosonic--type $S$--matrix with the aim 
to clarify the nature of its pathological behaviour, in 
particular at its short--distance scales. The TBA approach 
has shown that the difficulties consist in an instability 
of the vacuum state of the theory which occurs at a certain 
distance or energy scale, i.e. in the failure of the usual 
saddle point approach to minimise the free--energy. This quantity 
in fact for $R < R_c$ becomes in our case unbounded from below. 
Even in the absence of the above method, the peculiar physical 
behaviour of the QFT associated to a bosonic $S$--matrix may be 
inferred by the computation of its Form Factors. We have seen 
in fact that even with some stringent constraints imposed on 
the matrix elements they are intrinsically largely undetermined. 
Moreover, the spectral series which originate from them are 
generically divergent. 

Even after the thorough analysis of this paper, one 
could be still surprised that an innocent $-1$ in front of an 
$S$--matrix is the source of all the pathological features 
of the QFT associated to such an $S$--matrix. The fact is 
that the $S$--matrix, even for the integrable models where 
it takes a particularly simple form, is still the final 
result of an infinite resummation of all microscopic 
processes dictated by an Hamiltonian. Therefore while it 
is in general relatively simple to decide whether or not an 
Hamiltonian may give rise to a consistent Quantum Theory -- 
in terms of stability of the vacuum and the relative excitations 
thereof --  (once again, the simple example of footnote 5 is 
particularly instructive), the model analysed in this paper 
shows on the contrary that it may be in general ``algorithmically'' 
difficult to trace all this information back and to infer 
the consistency of a theory starting from the end, i.e. from 
the knowledge alone of the $S$--matrix. 

Several questions arise in relation to the observations of 
the previous sections. For instance, it would be interesting 
to analyse the QFT associated to more general CDD factors and 
characterise those which have a smooth interpolation from the 
short to the large distance scales, i.e. those ones which identify 
consistent QFT. Moreover, actions which are initially obtained by a 
perturbation of irrelevant fields contain an infinite series of 
corrections due to the higher dimension operators and hence 
they are generally non--local. In this respect, it would be 
useful to develop a powerful method for determining in an 
efficient way all the coupling constants of these operators 
and to have a criterion to identify those non--local theories 
which give rise to integrable models. 

\vspace{18mm} {\em Acknowledgements}. We wish to thank E. Corrigan, 
G. Delfino, M. Fabrizio, A. Parola, A. Schwimmer and Al.B. Zamolodchikov 
for discussions.

\newpage 

\appendix

\appsection

Let us consider the calculation of the free--energy of a bosonic 
type $S$--matrix with a kernel $\varphi(\beta)$ given by 
\EQ
\varphi(\beta) = 2\pi \delta(\beta) \,\,\,,
\label{localkernel}
\EN 
i.e. with a phase--shift defined by 
\EQ
-i \ln S(\beta) = \left\{
\begin{array}{ll}
-\pi & \mbox{if $\beta < 0$} \\
0 & \mbox{if $\beta =0$} \\
+\pi & \mbox{if $\beta > 0$} 
\end{array}
\right.
\label{phaseshift}
\EN 
In virtue of the locality of the kernel (\ref{localkernel}), 
the total density of levels $\rho(\beta)$ is simply related 
to the root density of levels by 
\EQ
\rho(\beta) = a(\beta) + \rho^{(r)}(\beta) \,\,\,, 
\label{algebraic}
\EN 
where 
\EQ
a(\beta) \equiv \frac{m}{2\pi} \cosh \beta \,\,\,.
\label{aa}
\EN 
The functional of the free--energy (at zero chemical potential) is 
given by 
\EQ
f(\rho,\rho^{(r)}) = {\cal E}(\rho^{(r)}) - \frac{1}{R} 
{\cal S}(\rho,\rho^{(r)})  
\label{functional1}
\EN 
where  
\EQ
{\cal E}(\rho^{(r)}) = \int m \cosh\beta \,\rho^{(r)}(\beta) \,
d\beta \,\,\,,
\label{hamiltonian1}
\EN 
and 
\EQ
{\cal S}(\rho,\rho^{(r)}) = \int d\beta 
\left[(\rho + \rho^{(r)}) \log(\rho + \rho^{(r)}) - 
\rho \log \rho - \rho^{(r)} \log \rho^{(r)} \right] \,\,\,. 
\label{entropy1}
\EN 
By plugging the expression (\ref{algebraic}) for $\rho(\beta)$ 
into eq.(\ref{functional1}) and minimizing with respect to the 
distribution $\rho^{(r)}(\beta)$, we have the following condition 
for the density $\rho^{(r)}$ which minimizes the functional
(\ref{functional1}) 
\EQ
\frac{
\left(a  + 2 \rho^{(r)}\right)^2}
{\rho^{(r)}\,\left(a  + 2 \rho^{(r)}\right)^2}
\,=\,{\cal C}
\,\,\,, 
\label{minmm}
\EN 
where 
\EQ
{\cal C} = e^{m R \cosh\beta} \,\,\,.
\EN 
For any value of $a$, the left hand side of eq.(\ref{minmm}) 
as a function of the positive values of $\rho^{(r)}$ is always 
larger than 4. Hence a positive real solution does not exist for 
{\em all} values of $R$ but only for those for which it is
verified the condition ${\cal C} \geq 4$. In this case the 
solution is given by 
\EQ
\rho^{(r)}(\beta) = 
\frac{a}{2} \left[\sqrt{\frac{{\cal C}}{{\cal C}-4}} -1 \right]
\,\,\,.
\EN 
Correspondingly 
\EQ
\rho(\beta) = 
\frac{a}{2} \left[\sqrt{\frac{{\cal C}}{{\cal C}-4}} +1 \right]
\,\,\,.
\EN 
Substituing these expressions into (\ref{functional1}), the value 
of the functional at its minimum is given by 
\EQ
\hat F(R) = m \int \cosh\beta \,
\ln\left[\frac{1}{2} \left(1 + \sqrt{1 - 4 e^{-m r \cosh\beta}}\right)
\right] \,\,\,.
\EN 
Hence, within the saddle point solution of the TBA equation we have 
\EQ
E_0(R) = \frac{2\pi}{R} G(m R) \,\,\,,
\EN 
where 
\EQ
G(x) = \frac{1}{\pi} \int_1^{\infty} \frac{dt \,t}{\sqrt{t^2-1}} \,
\ln\left[\frac{1}{2} \left(1+ \sqrt{1- 4 e^{-x t}}\right)\right] \,\,\,.
\EN 
This expression becomes complex for $R \leq R_c = \frac{\ln 4}{m}$. 
$R_c$ is the value where the first instability of the functional 
(\ref{functional1}) shows up, due to the distribution modes 
$\rho^{(0)}(\beta)$ and $\rho(\beta)$ computed at $\beta =0$ (for $R< R_c$ 
additional instabilities are induced by other modes). 
To see this, first of all notice that, in view of the 
algebraic equation (\ref{algebraic}), we have a complete decoupling of the 
contributions due to different $\beta$ in the free energy. Let us 
consider then the term $f^0$ in (\ref{functional1}) due to the two 
distributions computed at $\beta=0$. By using eq.(\ref{algebraic}), 
with the notation $\tilde\rho \equiv \rho^{(r)}(0)$ and $a_0\equiv 
\frac{m}{2\pi}$ we have 
\EQ
f^0(\tilde\rho,R) = \tilde\rho - \frac{1}{R} 
\left[
(a_0 + 2\tilde\rho) 
\ln(a_0 + 2\tilde\rho) 
- (a_0 + \tilde\rho) \ln(a_0+\tilde\rho) - 
\tilde\rho \ln\tilde\rho\right] \,\,\,.
\label{zeromode}
\EN 
This expression, as a function of $\tilde\rho$, is plotted in 
Figure 6 for $R > R_c$, $R=R_c$ and $R < R_c$ respectively. 
For $R > R_c$ $f^0$ admits a minimum for a positive real value of 
$\tilde\rho$ and the function is asymptotically positive. Hence 
for these range of $R$ the saddle--point approximation results 
valid. At $R=R_c$ the minimum has moved at infinity although the value 
of the function $f^0$ has remained finite. For $R < R_c$ the 
function $f^0$ does not have any longer a minimum and its 
values are unbounded from below, causing an instability in the 
corresponding free energy.

\newpage

\hs

{\bf Figure Captions}

\vspace{5mm}

\begin{description}
\item [Figure 1]. Renormalization Group flow which passes by a 
fixed point along the directions of the irrelevant and relevant 
fields $\eta$ and $\Phi$. 
\item [Figure 2.a]. Plots of the central charge $c(m R) = -
\frac{6 R}{\pi} F(m R)$ for the mShG model at a particular 
value of $a$, for the thermal Ising model and for a free 
bosonic model. Below $mR_c$ ($mR_c \sim 1$ in the figure), the
central charge of the mShG model assumes complex values. 
\item [Figure 3]. Profiles of the pseudo--energy $\epsilon$ 
versus $\beta$ for different values of R. $R_c$ corresponds 
to the value when this function hits the origin. 
\item [Figure 4]. Two--particle contribution to the $c$--theorem 
versus the parameter $a$ of the $S$--matrix. 
\item [Figure 5]. Graph of $\frac{{\cal F}(a)}{2 \sin\pi a}$ 
versus $a$. 
\item [Figure 6.a] 
The free--energy as a function of $\rho^{(r)}(0)$ for $R > R_c$, 
when there is a minimum. 
\item [Figure 6.b] The free--energy as a function of 
$\rho^{(r)}(0)$ for $R = R_c$. The minimum has moved at infinity. 
\item [Figure 6.c]. The free--energy as a function of 
$\rho^{(r)}(0)$ for $R < R_c$ when there is no longer a minimum.  

\end{description}

\pagestyle{empty}
\newpage

\begin{figure}
\null\vskip 3cm 
\centerline{
\psfig{figure=rgflow.ps}}
\vspace{1cm}
\begin{center}
{\bf \large{Figure 1}}
\end{center}
\end{figure}
\pagestyle{empty}

\newpage

\pagestyle{empty}

\begin{figure}
\null\vskip 3cm 
\hspace{1cm} \psfig{figure=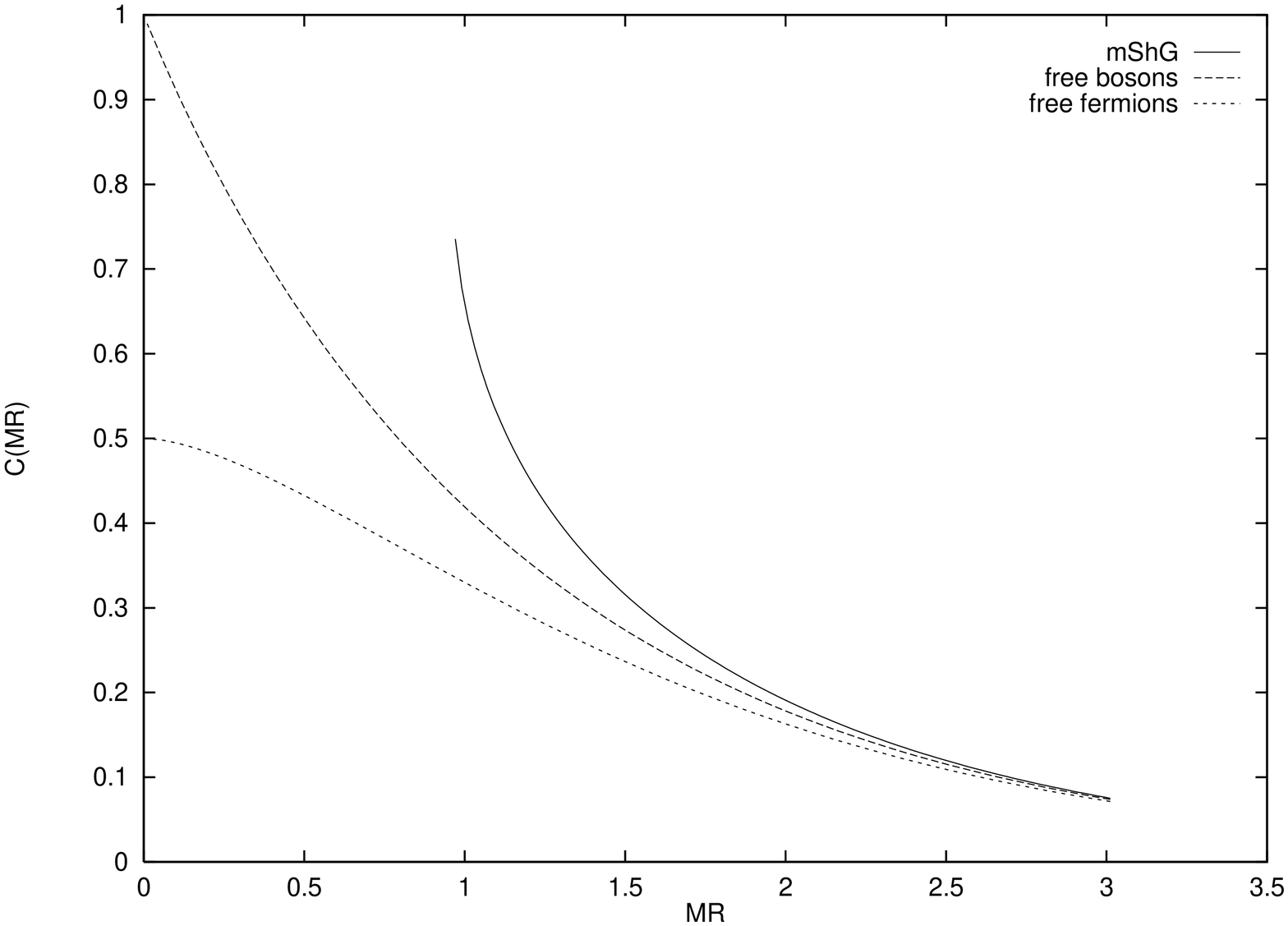,height=10cm,width=15cm,angle=-90} 
\vspace{1cm}
\begin{center}
{\bf \large{Figure 2}}
\end{center}
\end{figure}

\pagestyle{empty}

\newpage

\begin{figure}
\null\vskip 3cm 
\hspace{1cm}
\psfig{figure=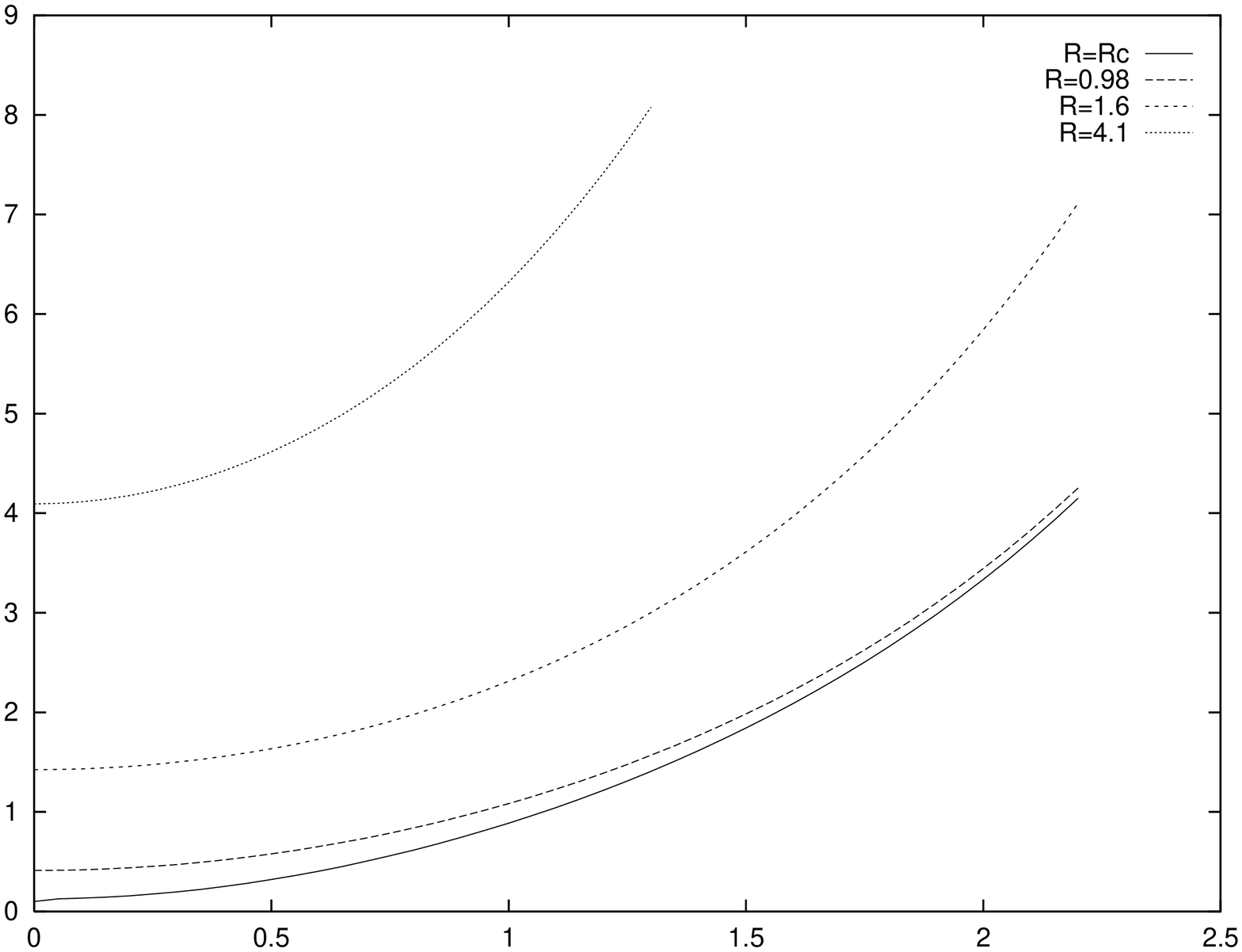,height=10cm,width=15cm,angle=-90} 
\vspace{1cm}
\begin{center}
{\bf \large{Figure 3}}
\end{center}
\end{figure}

\pagestyle{empty}
\newpage
\begin{figure}
\centerline{
\psfig{figure=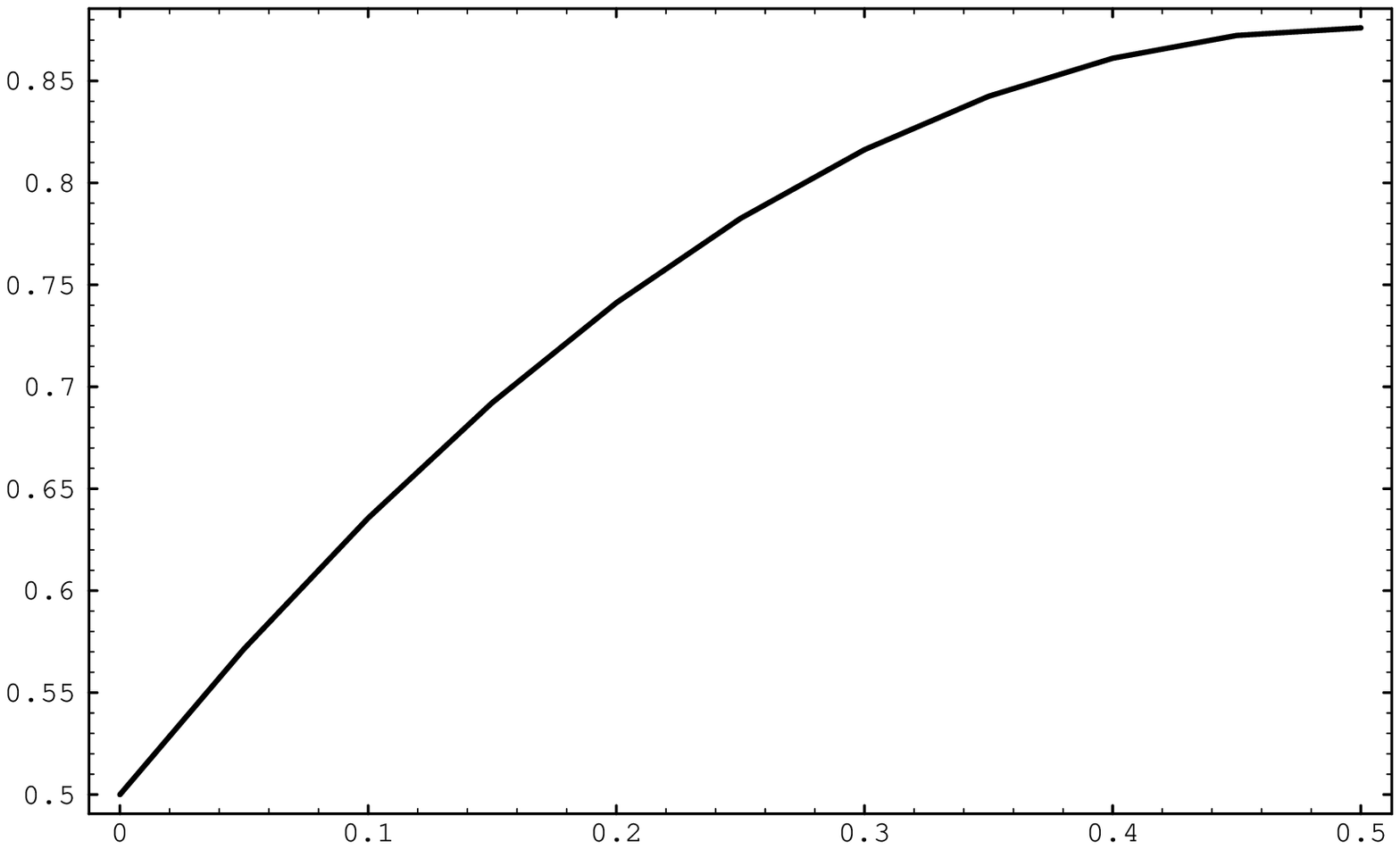}}
\vspace{-5mm}
\begin{center}
{\bf \large{Figure 4}}
\end{center}
\end{figure}

\newpage
\pagestyle{empty}

\pagestyle{empty}
\newpage
\begin{figure}
\centerline{
\psfig{figure=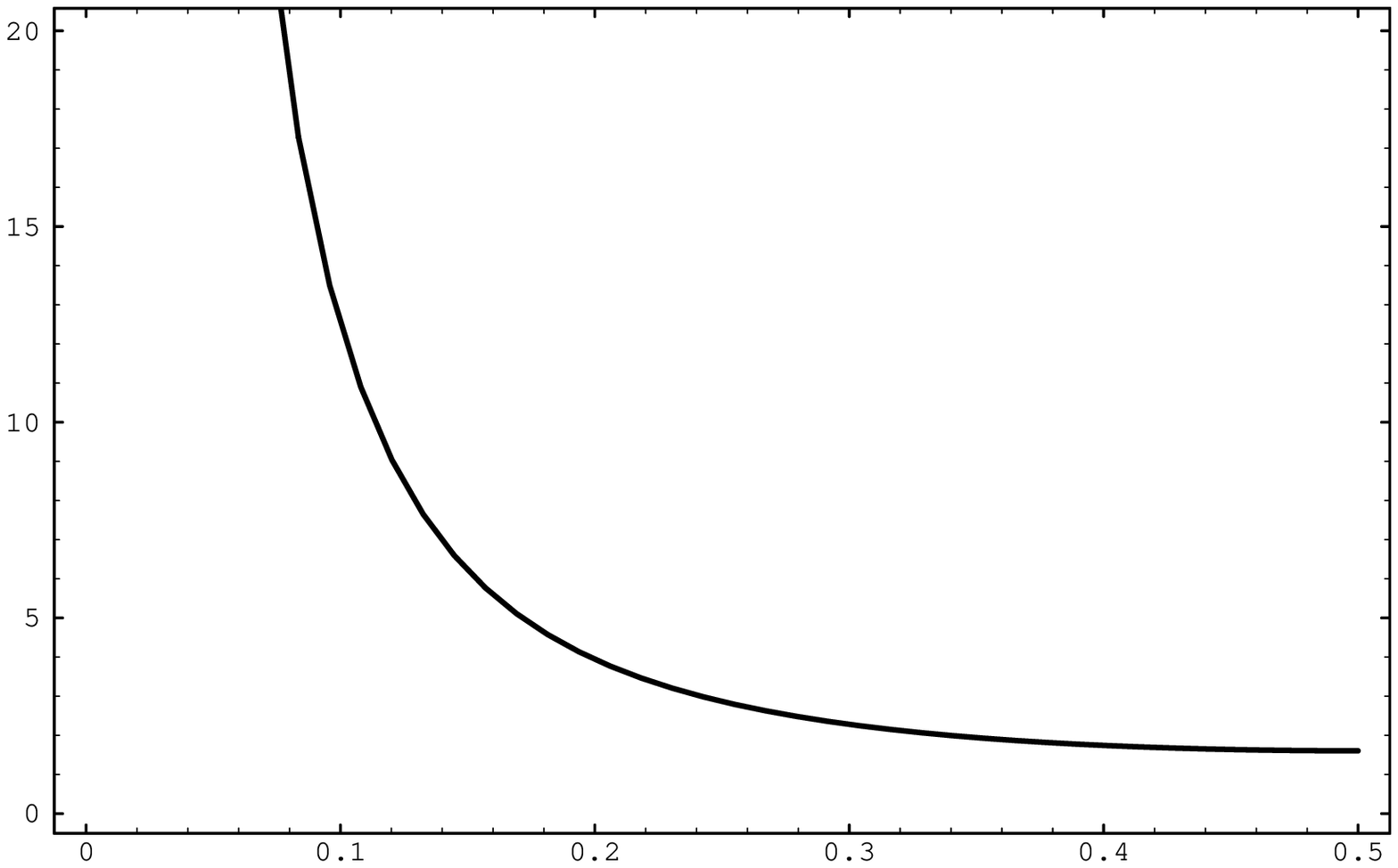}}
\vspace{-5mm}
\begin{center}
{\bf \large{Figure 5}}
\end{center}
\end{figure}

\newpage
\pagestyle{empty}
\begin{figure}
\null\vskip -15mm 
\centerline{
\psfig{figure=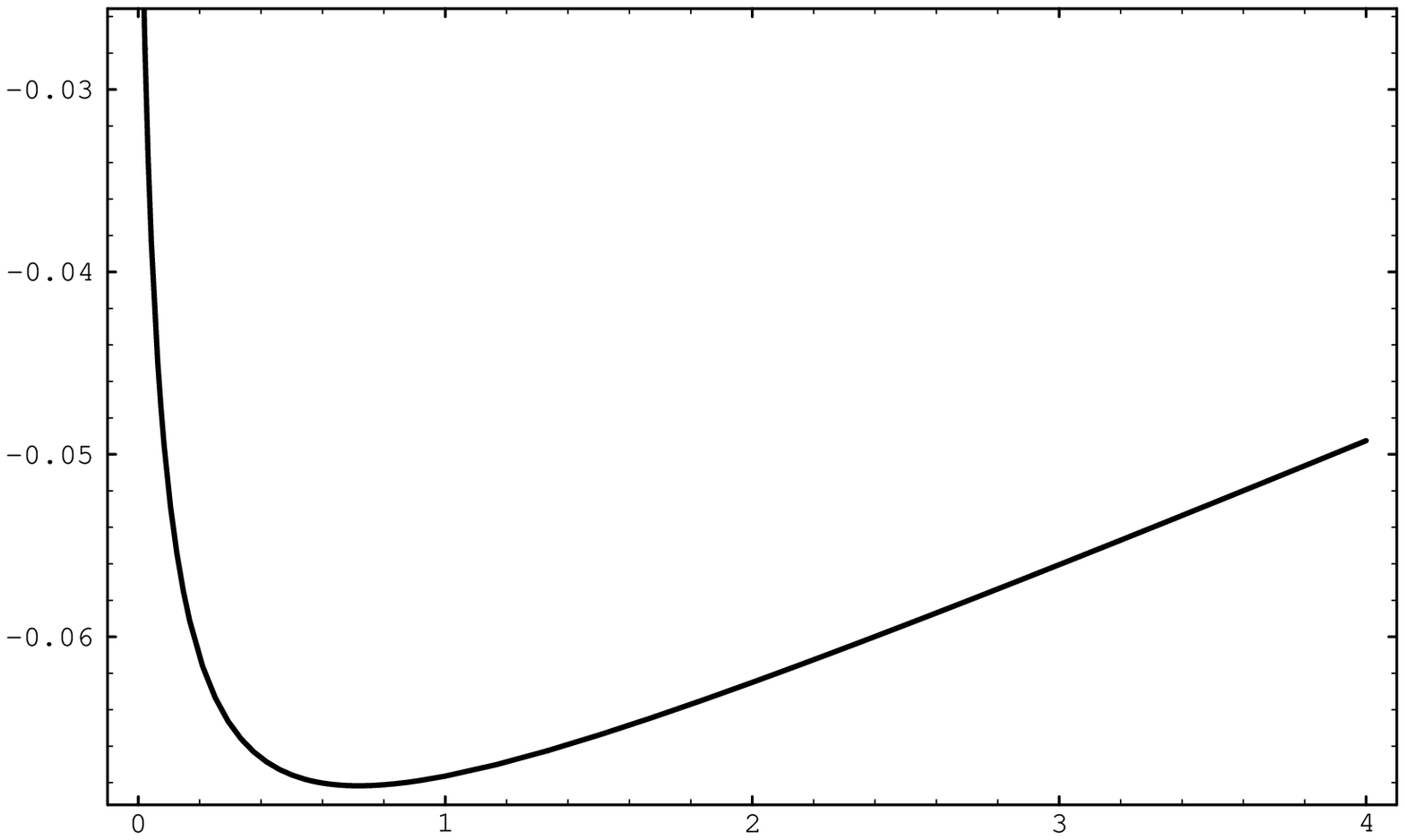}}
\vspace{-5mm}
\begin{center}
{\bf \large{Figure 6.a}}
\end{center}
\end{figure}

\newpage
\pagestyle{empty}

\begin{figure}
\null\vskip -15mm 
\centerline{
\psfig{figure=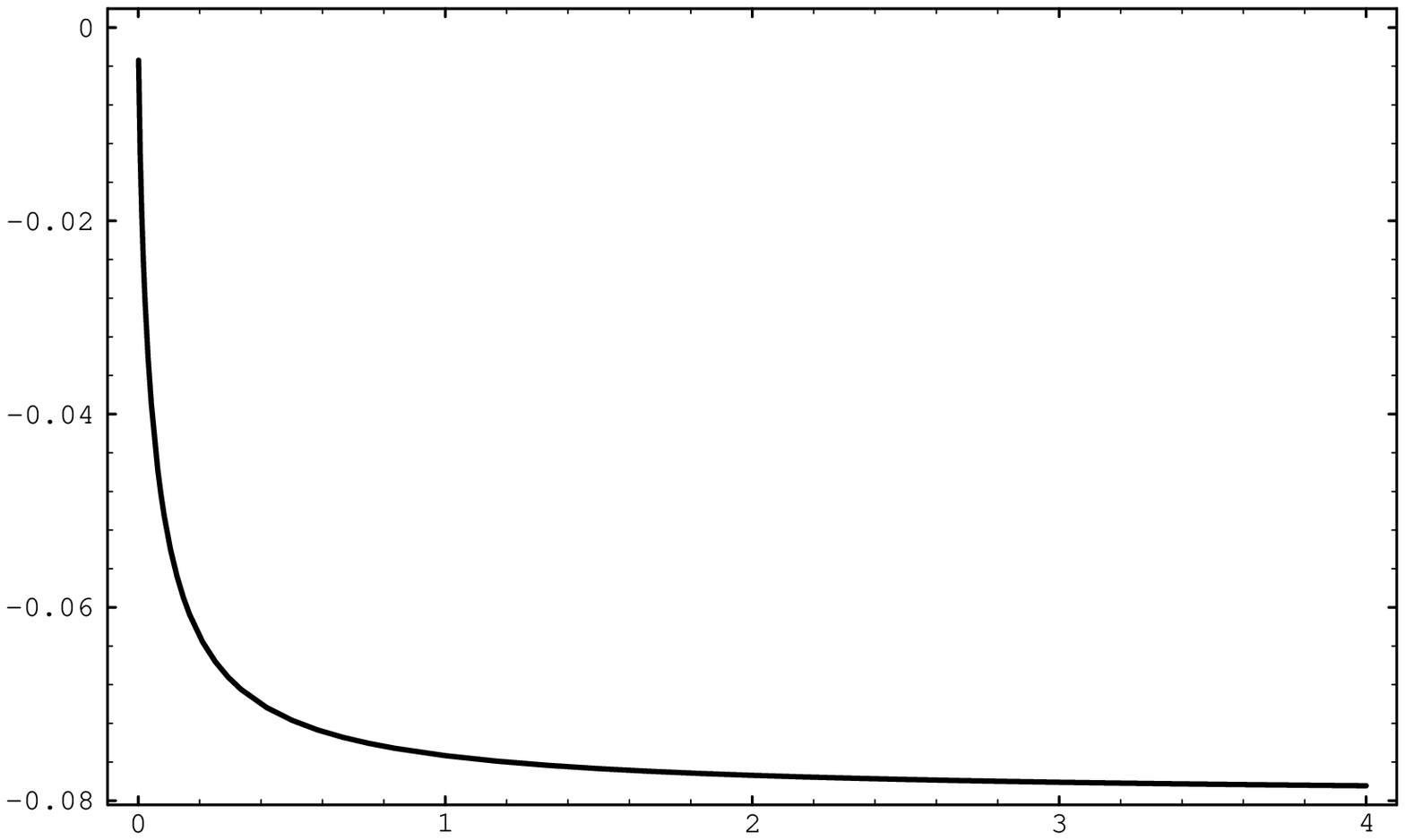}}
\vspace{-5mm}
\begin{center}
{\bf \large{Figure 6.b}}
\end{center}
\end{figure}

\newpage
\pagestyle{empty}

\begin{figure}
\null\vskip -15mm 
\centerline{
\psfig{figure=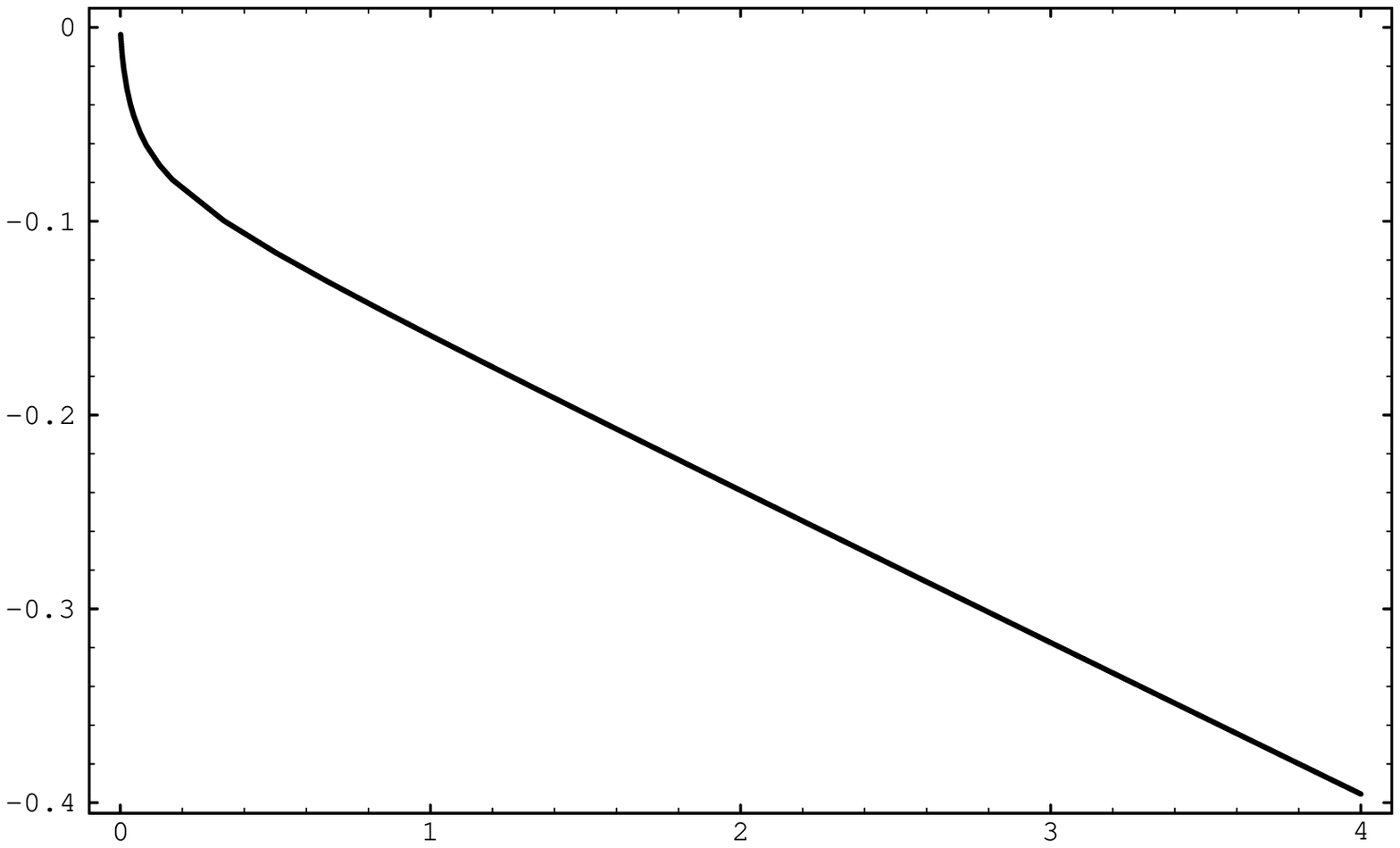}}
\vspace{-5mm}
\begin{center}
{\bf \large{Figure 6.c}}
\end{center}
\end{figure}

\end{document}